\documentclass[<options>]{elsarticle}

\usepackage{mathtools}
\usepackage{xcolor}
\usepackage{hyperref}
\hypersetup{
    colorlinks=true,
    linkcolor=blue,
    filecolor=magenta,      
    urlcolor=cyan,
    pdftitle={Overleaf Example},
    pdfpagemode=FullScreen,
    }

\begin{document}
\title{Analysis of practical fractional vortex beams}

\author[1,2]{Eduardo Peters}
\author[1,2,*]{Gustavo Funes}
\author[3]{L. Mart\'inez-Le\'on}
\author[3]{Enrique Tajahuerce}

\address[1]{Universidad de los Andes, Chile.}
\address[2]{Millennium Institute for Research in Optics (MIRO), Chile.}
\address[3]{GROC-UJI, Institute of New Imaging Technologies (INIT), Universitat Jaume I, 12071 Castelló, Spain.}
\address[*]{Corresponding author: gfunes@miuandes.cl}

\begin{abstract}
The subject of calculating the topological charge (TC) or vortex strength of optical vortices has generated divided opinions among scientists. This is due to the fact that proper analytical results are hard to support from the experimental point of view, leading to different results and conclusions. In this work we will present numerical data that shows the limits of TC measurements for practical fractional vortices and the possible challenges that high order measurements may pose. By analyzing the far field phase and the behavior of the transitions we have shown that they follow specific curves that depend not only on the TC but also on the beam waist. This leads us to present a new ``strength staircase'' for practical vortices. Our aim is to give some insight in practical scenarios that have not been taken into account in previous results.
\end{abstract}


\maketitle

\section{Introduction}

Singular beams have been studied thoroughly since the first works of Allen \textit{et al.} back in 1992 \cite{paper:Allen1992}. After that, there has been a lot of research and applications in this area. A particular kind of singular beams are the so called Fractional Vortex Beams, (FVBs) first observed experimentally by Beijersbergen \textit{et al.} \cite{paper:Beijersbergen1994}. FVBs, or more correctly called non-integer beams, are a type of singular beams that possess a non-integer singularity in opposition to the regular Laguerre-Gaussian beams. FVBs have some quaint characteristics that lead to debate and some disagreements among researchers. One of the most interesting feature is its topological charge or beam strength and how to measure it, in the near and far field. According to references \cite{paper:Berry2004,paper:Basistiy2004} the fractional topological charge evolves from fractional (in the near field) to an integer (in the far field). 
In the work of Leach \cite{paper:Leach2004}, they analyze the FVB phase in the near field and measure its topological charge. Then Lee \textit{et al.} \cite{paper:WLee2004} study FVBs using an interferometric technique and discover that there is a dynamic in the transition from one integer vortex to another. This transition, that was often called ``birth of a vortex'' has led to the discussion of the actual value of the TC between two integer states. As well as the regular singular beams, the applications of FVBs are diverse. Most of them are similar to those of regular singular beams. They include particle trapping \cite{paper:STao2005}, perfect FVBs \cite{paper:Tkachenko2017}, analysis of optical forces \cite{paper:Bing2021}, and electromagnetic vortex beams for imaging \cite{paper:Liu2021}. For the particular case of communications, vortices have been used as symbols \cite{paper:shuhuili2017} or in multiplexing scenarios \cite{paper:Duodeng2021} \cite{paper:Kuang2020}. Specifically in communications, the ability of detecting the vortex strength is a key factor. For that reason and for scientific purposes, great effort has been dedicated to measuring the fractional TC. Moreover, in references \cite{paper:Berry2004}, \cite{paper:Alperin2016}, \cite{paper:Gbur2016} and more recently in \cite{paper:Matta2020} the authors thoroughly analyze the phase structure of FBVs and the transition from near to far field. In spite of this effort, some discrepancies among scientists arise related to the determination of the vortex strength in the near and far field

In this work we show by numerical simulations the limits of TC measurements for realistic fractional vortices and some possible problems arising when higher-order measurements are performed. We prove that the behavior of the vortices during the continuous transition between two integer values of the TC follow specific curves that depend not only on the value of the TC but also on the beam waist of the light beam. Our study may help to give some insight in the results obtained in practical situations. 

\begin{figure}[hb!]
\centering\includegraphics[width=0.9\linewidth]{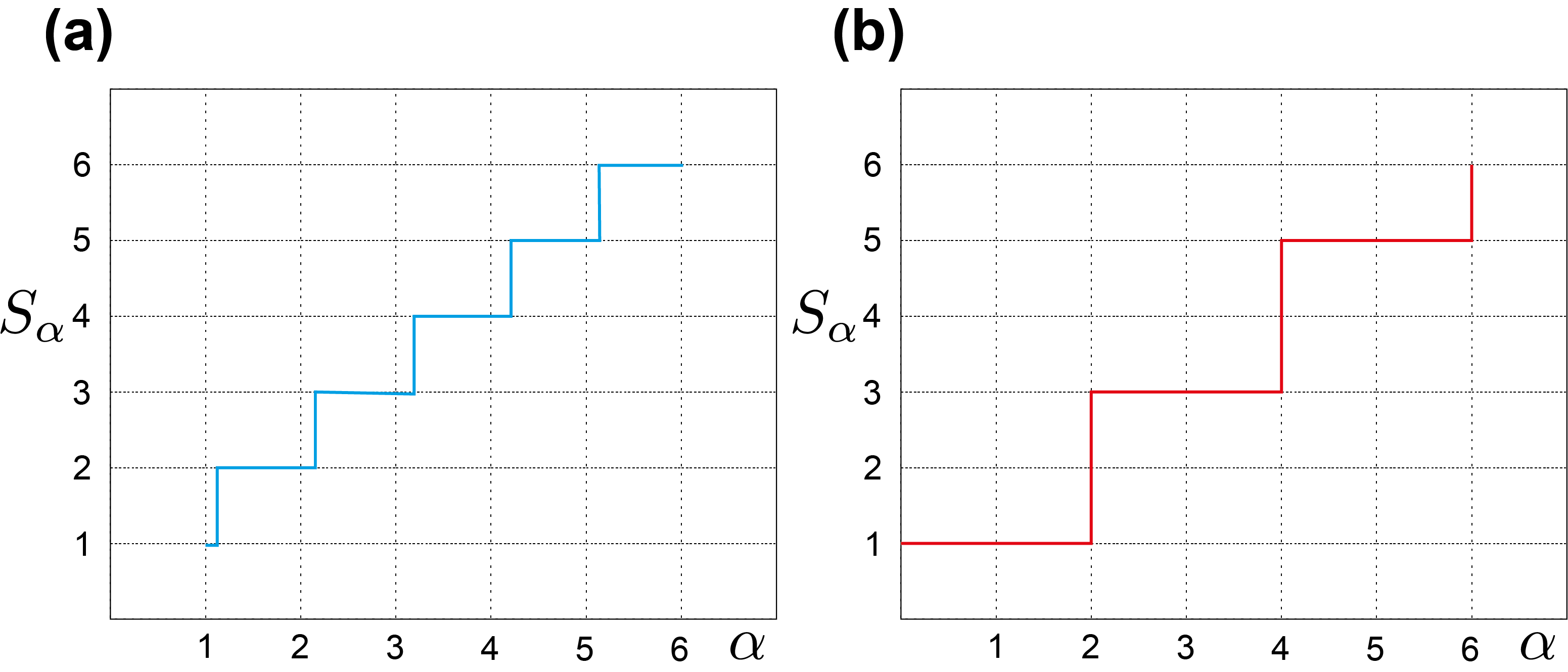}
\caption{Representation of the strength staircase according to (a) Jesus-Silva \textit{et al.} \cite{paper:Jesus-Silva2012} and (b) Wen \textit{et al.} \cite{paper:Wen2019}.}
\label{fig0000}
\end{figure}

\section{The strength staircase}

As mentioned previously, the fractional TC evolves from fractional in the near field to an integer in the far field. When the TC changes continuously from one integer value to the next one, taking decimal values, the vortex strength should be an integer value. In principle, the curve representing the vortex strength as a function of the TC should be a staircase, as shown in Figure \ref{fig0000}. Nevertheless, the form of the staircase is the key of the discrepancy. For example, when studying near field measurements, as performed in \cite{paper:Alperin2016,paper:Duo2019}, they show what can be called a ``soft staircase'' when passing from one integer to another, in opposition with the numerical predictions of Berry \cite{paper:Berry2004}. It is also important to mention other works that measure vortex strength in far field and report a linear dependence \cite{paper:Hosseini2020,paper:Guoxuan2021}. This clearly is a controversial issue because the reported TC depends on the measurement technique and whether the vortex is observed in the near or far field. It is known that the dynamics of the fractional vortex beams from near to far field present a string of alternating vortices \cite{paper:Berry2004,paper:Basistiy2004, paper:Alperin2016, paper:Gbur2016} that disappear as the vortex reaches the far field, but this is not the whole picture. Some explanation of this dichotomy regarding whether the FVB evolves to an integer strength in far field or not is provided by Alperin and Siemens \cite{paper:Alperin2017} but the evidence is not conclusive.

The dynamics of FVBs passing from one integer to another was thoroughly investigated recently by Wen \cite{paper:Wen2019} using numerical results validated by experiments. Nevertheless, some aspects of this vortex transition are not well understood yet. Wen \textit{et al.} have discovered that the ``staircase'' of vortex strength vs initial topological charge $\alpha$ is always an odd integer in the far field. Their results pose a discrepancy with previous works like \cite{paper:Berry2004,paper:Gbur2016, paper:Jesus-Silva2012} that obtain the ``TC staircase'' but with different values.
%and finally led to further study by Kotlyar \textit{et al.}. \cite{paper:Kotlyar2020a,paper:Kotlyar2020b} who developed an analytical model for vortex propagation and they also supported their discoveries with numerical simulations and experiments.  

One way to evaluate the vortex strength, $S_{\alpha}$, for different values of the predetermined TC, $\alpha$, is by using the following equation \cite{paper:Berry2004}:

\begin{equation}
    S_{\alpha}=\lim_{R\to\infty} \frac{1}{2\pi}\int^{2 \pi}_0 d\varphi \frac{\partial}{\partial \varphi} arg\, U (\textbf{r})
    \label{eq:000}
\end{equation}

where \textit{U}$(\textbf{r})=A(\textbf{r})e^{i \psi (\textbf{r})}$ denotes the light field, and R is the radius of a curve surrounding the optical axis. From a numerical point of view, it is possible to calculate \textit{S} for the transition from one integer to another with arbitrary precision. The result coincides with the number of integer vortices present in the beam. Nevertheless, from a practical point of view this vortex strength equation suffers from the physical limitations of the waist of the Gaussian beam that creates it. 

\begin{figure}[ht!]
\centering\includegraphics[width=1\linewidth]{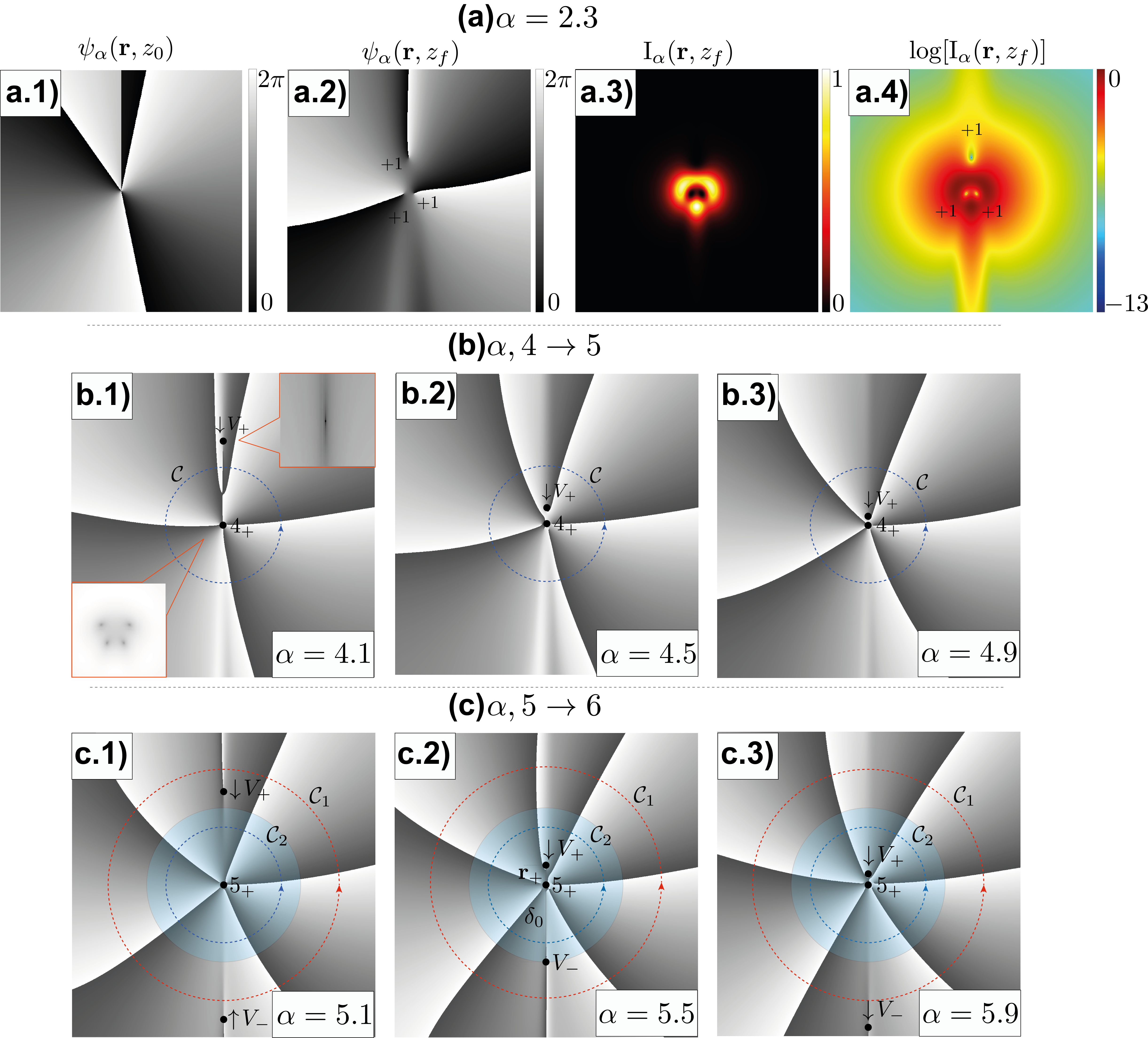}
\caption{(a) Fractional vortex beam $\alpha=2.3$. (a.1) Initial phase profile $\psi_{\alpha}(\mathbf{r},z_{0})$. (a.2) Far-field phase profile $\psi_{\alpha}(\mathbf{r},z_{f})$. (a.3) Normalized far-field intensity profile $I_{\alpha}(\mathbf{r},z_{f})$. (a.4) Far-field intensity profile (logarithmic scale) $\log[I_{\alpha}(\mathbf{r},z_{f})]$. (b) Far-field phase profiles samples for transition $\alpha, 4 \to 5$. (b.1) Vortex dynamics for $\alpha=4.1$. Top zoom: $+1$ vortex moving towards the nucleus. Bottom zoom: nucleus $+4$. (b.2) Vortex dynamics for (b.2) $\alpha=4.5$ and (b.3) $\alpha=4.9$. (c) Far-field phase profiles samples for transition $\alpha, 5 \to 6$. Vortex dynamics for (c.1) $\alpha=5.1$, (c.2) $\alpha=5.5$ and (c.3) $\alpha=5.9$.}
\label{fig1aa}
\end{figure}

Lets see the form of the intensity and phase of FVBs to understand the dynamics of the transition; Figure \ref{fig1aa}(a) shows the particular case $\alpha=2.3$ from the transition $2 \to 3$. The distribution of three unit vortices ($+1$) in the far field can be observed in the phase $\psi_{\alpha}$ (Fig. \ref{fig1aa}(a.2)). Figure \ref{fig1aa}(a.3) shows the intensity $I_{\alpha}$ in the far field revealing the lack of circular symmetry in comparison with non fractional (integer $\alpha$) vortex beams. The logarithm of the intensity allows to identify the local minima to relate them to the position of the unit vortices. 

When observing Figure \ref{fig1aa}(b) one can notice that in general there are two singularity zones in an even to odd transition while in an odd to even transition there are three. In the first one (Fig.\ref{fig1aa}(b)) the nucleus has the greater TC (except for transition $0\rightarrow 1$) while  the other region is conformed by a unitary vortex that we call the \textit{resident}. This is because this vortex appears from infinity and enters the nucleus. Similarly in the second type (Fig.\ref{fig1aa}(c)) there is one \textit{resident} vortex and another vortex of opposite charge that we call the \textit{tourist}. This last one reaches certain minimum point near the nucleus and then leaves. For this reason the TC measurement for the second type transitons always remain odd in the far field independent from the values of $\alpha$ involved. This is well documented in \cite{paper:Wen2019}.

The strength of practical FBVs (optical beams with finite width) depends not only on the initial charge $\alpha$, but also on the incident light source, therefore considering a Gaussian beam instead of an ideal plane wave produces a different output. This is of utmost importance because in many practical applications a Gaussian beam is incident upon a spatial light modulator (SLM) to generate an optical vortex.

Fist of all we will focus our attention in the previous ``staircases'' encountered in literature. According to Berry 2004 \cite{paper:Berry2004} the vortex strength in the near field make a unitary jump at half the way between two integers. For example when changing from $\alpha=1$ to 2 the TC is 1 up to $1.5$, then it turns to 2 and it jumps to 3 only when $\alpha$ reaches the value $2.5$. This result is also presented in \cite{paper:Gbur2016}. A very similar result is obtained by Jesus-Silva \textit{et al.} \cite{paper:Jesus-Silva2012} where the $\alpha$ value of the jumps move towards the lower integer (See Figure \ref{fig0000}(a)). That is, from $\alpha=2$ to 3 the TC jumps from 2 to 3 at approximately $\alpha=2.1$. And they verify it by counting the amount of unitary vortices within a fixed radius in the Fraunhoffer zone. According to \cite{paper:Wen2019} there are more unitary vortices involved than the previously reported by Jesus-Silva. Considering this, the staircase now makes two TC jumps at even values of $\alpha$ and remains odd around odd values of $\alpha$ as can be seen in Figure \ref{fig0000}(b).

\section{Analysis of practical FBVs}

In order to analyze rigorously the behaviour of the vortices and the vortex strength as a function of $\alpha$ for practical vortex beams, we will generate diverse FVBs focused on the Fraunhoffer region. For this task, we utilize a monochromatic Gaussian beam impinging normally onto an SLM with transmission function $\exp(i\alpha \theta)$ in the initial transverse plane $\mathbf{r}=(r,\theta)$. Then the initial field is $U_{\alpha}(r,\theta)=\exp(-r^2/\omega_{0}^2)\exp(i\alpha \theta)$, where $\omega_{0}$ is the beam waist radius and $\alpha$ is the topological charge TC.
Using the paraxial approximation, the field in the Fraunhofer diffraction zone ---aside from multiplicative phase factors--- is the two-dimensional Fourier transform of the initial field, evaluated at frequency $\rho/\lambda z$, i.e. $U_{\alpha}(\rho,\varphi)\propto \mathcal{F}[U_{\alpha}(r,\theta)]$ \cite{paper:Jesus-Silva2012}. Then, if a non-integer $\alpha$ is introduced in the original field, within the propagation it produces a step discontinuity in addition to the central singularity \cite{paper:Berry2004}.

For practical FVBs we will take the vortex strength as the signed sum of the vortices within a closed loop $\mathcal{C}$ including the $z$ axis as presented by Gbur \cite{paper:Gbur2016}, 
and can be determined by
\begin{equation}
S=\frac{1}{2\pi}\oint_{\mathcal{C}}\nabla \psi(\mathbf{r})\cdot d\mathbf{r},
\label{eq00}
\end{equation} 
where $\psi(\mathbf{r})$ is the wavefield phase and $\mathcal{C}$ has to be a simple contour in the mathematical sense. Both equations \ref{eq:000} and \ref{eq00} are equivalent if the phase is only dependent of $\varphi$. However, equation 2 is more convenient in calculating TC for practical FVBs, as it allows to take into account the radius of $\mathcal{C}$, that cannot have a limit to infinity for realistic situations. For most experimental cases the calculations are made by counting the singularities around the center \cite{paper:Jesus-Silva2012,paper:Wen2019}. This poses a problem in defining a radius for the contour. For singular vortex beams with integer topological charge $\alpha$, the vortex strength is independent of the contour due to the circular symmetry. However, if the topological charge is fractional the circular symmetry is broken due to edge dislocation and the vortex strength will no longer be independent of the chosen path. 

\begin{figure}[ht!]
\centering\includegraphics[width=\linewidth]{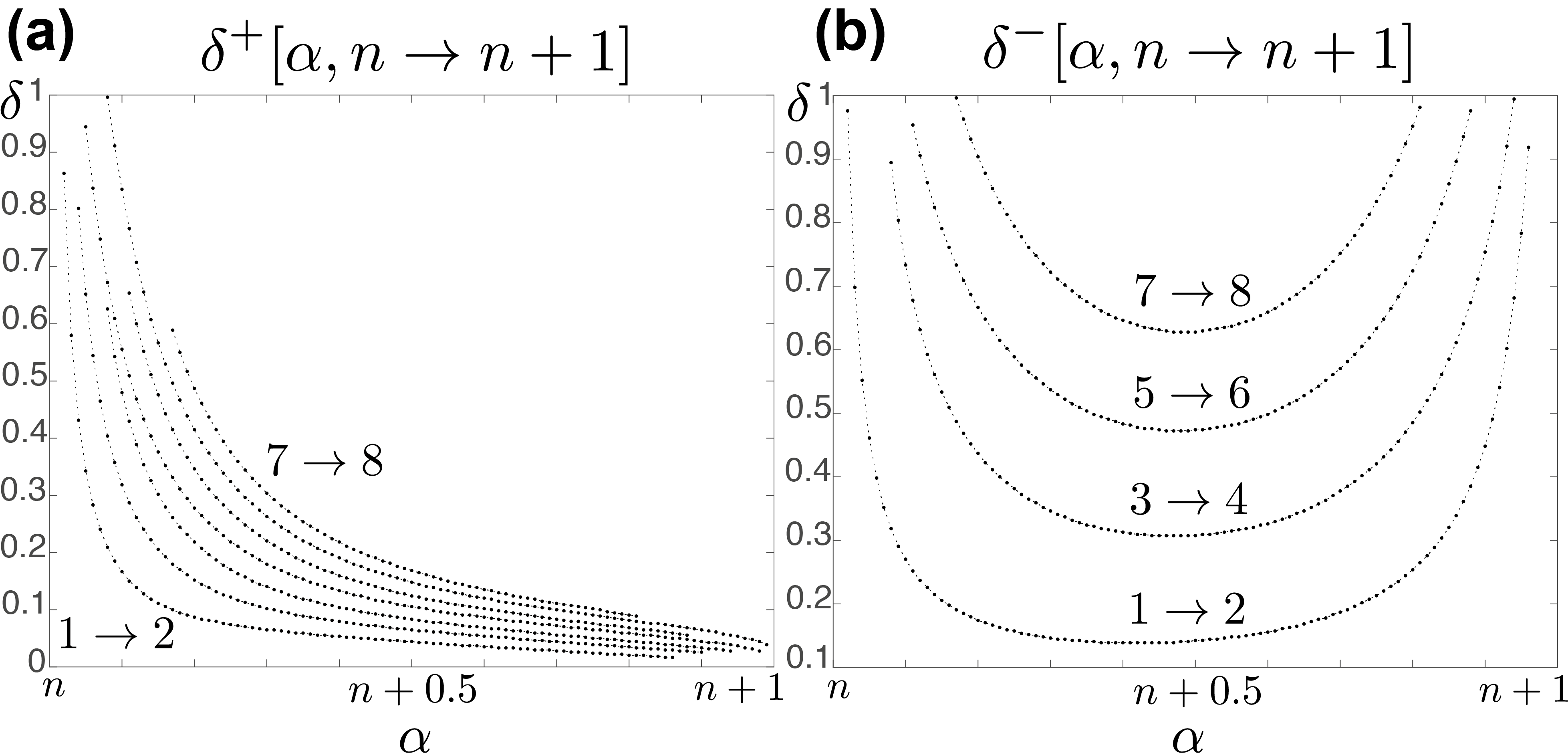}
\caption{Normalized distance $\delta=\delta(\alpha)$ measured from the position of the unit vortices to the center. (a) Distance $\delta^{+}$ for the \textit{resident} vortices in transitions from $1 \to 2$ to $7 \to 8$. (b) Distance $\delta^{-}$ for the \textit{tourist} vortices, present only in transitions starting with odd $n$, from $1 \to 2$ to $7 \to 8$.}
\label{fig2}
\end{figure}

Considering an integer jump directly from $\alpha=n$ to $\alpha=n+1$, there is circular symmetry, and the vortices are all in located in the center (nucleus) forming a single vortex with integer charge. For the case of a fractional transition there are two or three zones of singularities as mentioned earlier. 
We will define the value $\delta$ as the normalized distance\footnote{The normalization is obtained by dividing the real distance $d$ by the maximum possible distance for a fixed resolution} of the unit vortices from the center of the beam to their actual positions. Those distances can be obtained by detecting the vortices (see Fig.\ref{fig1aa}(a.4)) present over the intensity or phase distributions. Figure \ref{fig2} shows the distance $\delta$ as a function of the fractional topological charge, $\alpha$. All transitions have \textit{resident} type vortices, as can be seen in Fig. \ref{fig2}(a). We will define $\delta^{+}$ and $\delta^{-}$ as the distances from the center to the \textit{resident} and \textit{tourist} vortex respectively. 
The \textit{tourist} vortices only appear in odd to even transitions, as shown in \ref{fig2}(b), represented by $\delta^{-}$ for positive $\alpha$. This behavior is related to the parity of the integer topological charges at the borders of a transition. In this case, the \textit{tourist} vortex reaches a minimum distance $\delta_{0}$ and returns.

\begin{figure}[ht!]
\centering\includegraphics[width=1\linewidth]{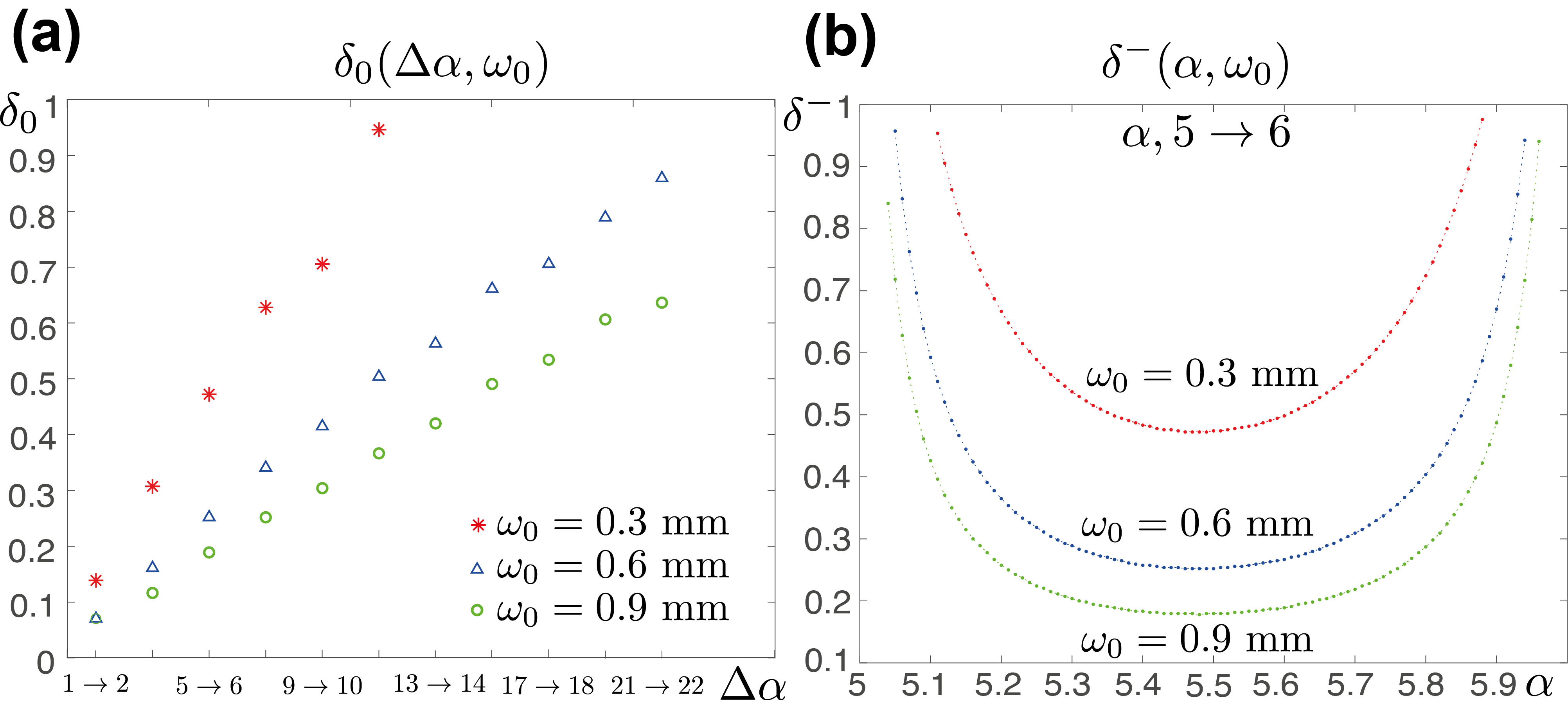}
\caption{(a) Normalized minimum distance $\delta_{0}$ as function of odd transitions $\Delta \alpha$ for different beam waist radius $\omega_{0}$. (b) Normalized distance $\delta^{-}$ for the \textit{tourist} vortex in transition $5 \to 6$, for different beam waist radius $\omega_{0}$.}
\label{fig1b}
\end{figure}

Lets return briefly to the example shown in Figure \ref{fig1aa}(c) where we can consider two contours, namely $\mathcal{C}_{1}$ and $\mathcal{C}_{2}$, where ---depending on $\alpha$--- $\mathcal{C}_{1}$ can count both vortices and $\mathcal{C}_{2}$ never encloses the \textit{tourist} vortex. The blue disk of radius $\delta_{0}$ represents the zone where this \textit{tourist} vortex never enters. Following Fig.\ref{fig1aa}(c.1)($\alpha=5.1$), $S=+6$ for $\mathcal{C}_{1}$ and $+5$ for $\mathcal{C}_{2}$. In Fig.\ref{fig1aa}(c.2)($\alpha=5.5$), $S=+5$ for $\mathcal{C}_{1}$ and $+6$ for $\mathcal{C}_{2}$. Finally, in Fig.\ref{fig1aa}(c.3)($\alpha=5.9$), $S=+6$ for $\mathcal{C}_{1}$ and $\mathcal{C}_{2}$.

For the odd to even FVBs transitions the dynamics includes both the \textit{resident} and \textit{tourist} vortices. It is important to mention two aspects of this: First, the tourist vortex turns around farther from the origin while the TC of the nucleus rises. We show this in Figure \ref{fig1b}(a), and for that reason the distance $\delta_{0}$ increases. Once this \textit{tourist} vortex is no longer visible it will not contribute to the beam strength posing a limit to the TC measurement. This limit depends on the simulation parameters or actual experimental setup, specifically on the beam waist $w_0$. The second thing to notice is that the \textit{tourist} vortex never gets closer to the center than the \textit{resident} vortex. Then, even though the numerical staircases \cite{paper:Wen2019} have uniform steps, in practical scenarios they will be modified beyond certain value of $\alpha$.
Moreover, as can be seen in Figure \ref{fig1b}, when the beam waist of the original Gaussian beam is modified, the slope of the $\delta_0$ curve changes, and with it, the behavior of the \textit{tourist}.

\begin{figure}[ht!]
\centering\includegraphics[width=\linewidth]{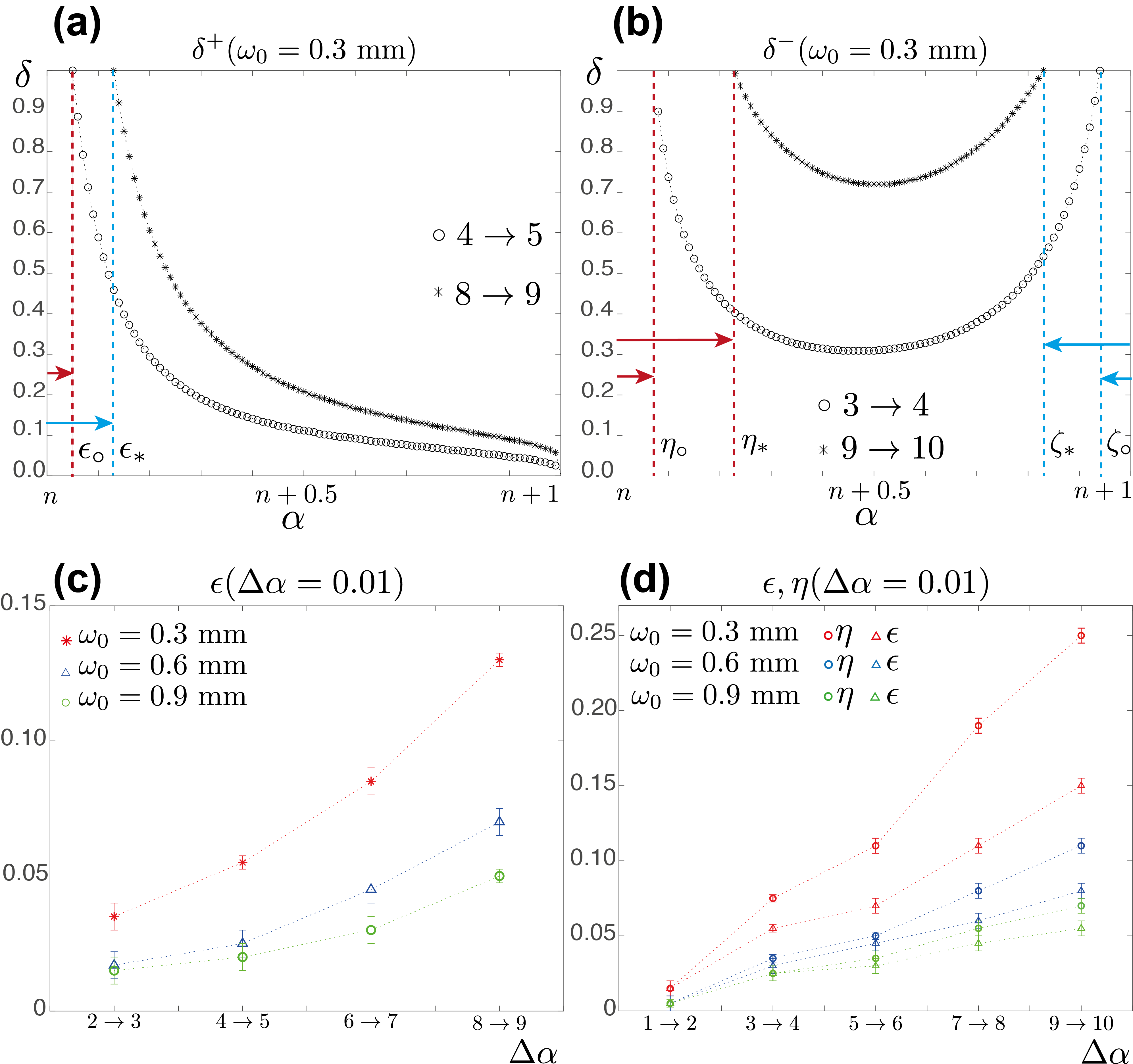}
\caption{(a)-(b) Normalized distance $\delta=\delta(\alpha, \omega_{0}=0.3\ \text{mm})$. (a) \textit{Resident} vortex for transitions $4\to 5$ (circle) and $8 \to 9$ (asterisk). (b) \textit{Tourist} vortex for transitions $3\to 4$ (circle) and $9 \to 10$ (asterisk). (c)-(d) Parameters $\epsilon$ and $\eta$ for step $\Delta \alpha =0.01$. (c) Parameter $\epsilon$ for even transitions with different beam waist $\omega_{0}$. (d) Parameters $\epsilon$ and $\eta$ for odd transitions with different beam waist $\omega_{0}$.}
\label{fig3a}
\end{figure}

\begin{figure}[ht!]
\centering\includegraphics[width=0.9\linewidth]{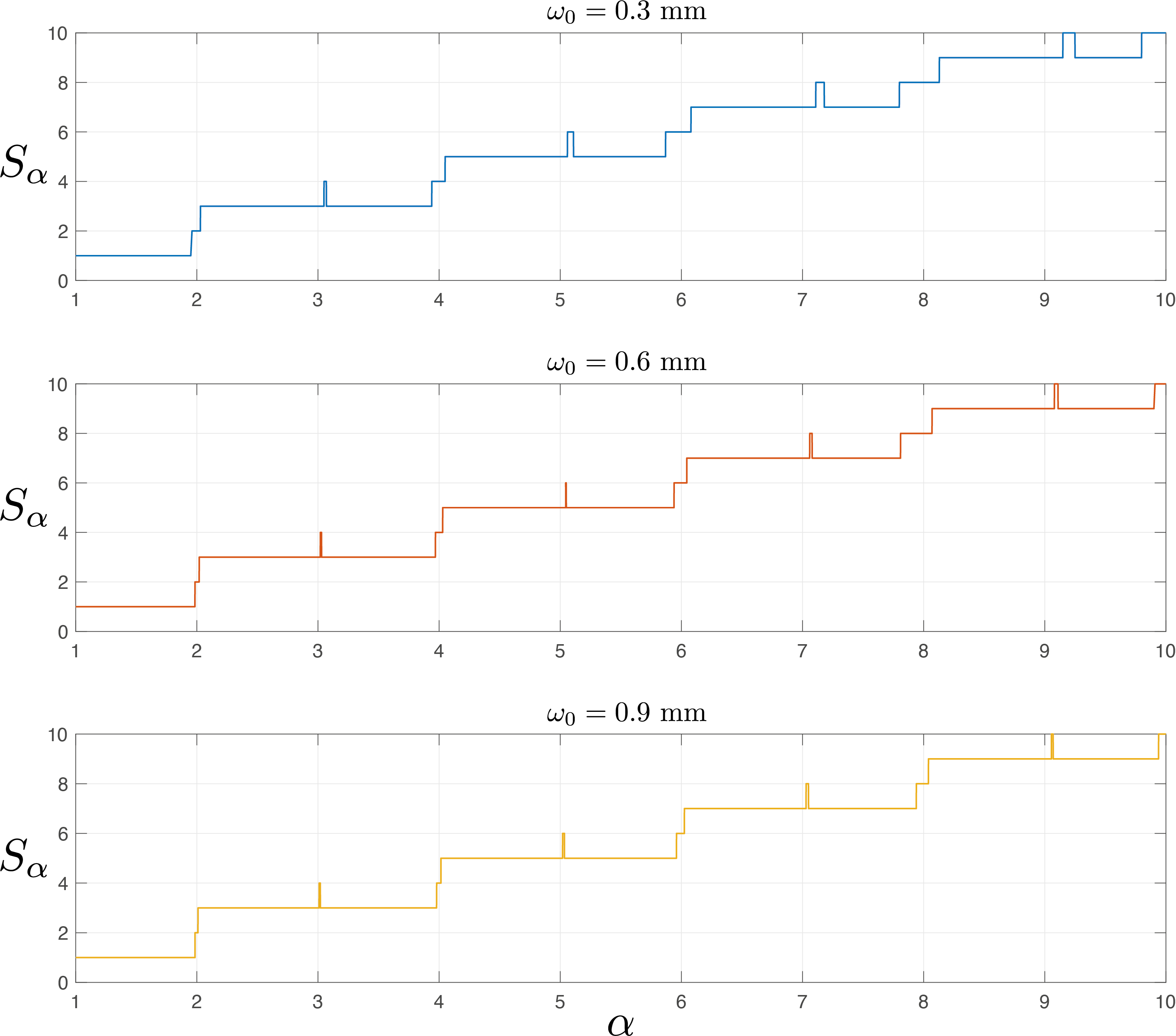}
\caption{The strength staircase for different beam waists based on the values of $\epsilon$, $\eta$ and $\zeta$}
\label{fig3b}
\end{figure}

Now taking into account that the curve $\mathcal{C}$ can be modified and that the beam waist $\omega_0$ is fixed, we can argue that the limit of the vortex extension is the radius of the curve. In this fashion, we can obtain the values in alpha at which the resident curve cuts the edge of the beam, $\varepsilon$, and the values at which the tourist cuts the same edge, $\eta$ and $\zeta$. Next, we analyze the values of $\varepsilon$, $\eta$ and $\zeta$ for different cases. Figure \ref{fig3a} (a) and (b) shows the graphical interpretation of the new parameters. We have estimated them for three different beam waists $\omega_0$ and transitions from 1 to 10. These estimations can be observed in Figure \ref{fig3a} (c) and (d) where we have separated transitions of the first kind (even to odd) from the second kind (odd to even). This is because the existence of the \textit{tourist} affect the behavior or the \textit{resident}, as can be seen in Figure \ref{fig3a}(d). We have omitted the parameter $\zeta$ since its behavior is similar to $\eta$. The evolution of the parameter $\epsilon$ was briefly mentioned in \cite{paper:Jesus-Silva2012} but never explained in detail. When comparing the values of $\eta$ and $\epsilon$ seems that the first is bigger than the second. This means that the \textit{resident} enters in the beam extension before the \textit{tourist} does. This is true at least for smaller beam waists or curves $\mathcal{C}$ with short radius. In a numerical scenario using equation \ref{eq:000} in the limit of infinite R, both curves converge. Now this can be problematic in real experiments since $\omega_0$ is finite, and can be different depending on the experimental setup. It also limits the properties of the beam that the scientist want to observe. That is, if $\omega_0$ is small, the extension of the beam in the far field is reduced but the resolution near the center is higher. On the contrary, if $\omega_0$ is large more of the external parts of the beam can be seen but paying the price of getting a lower resolution in the center. Continuing with this line of thought and considering the variation of $\delta_0$ as shown in Figure \ref{fig1b}, it is clear that the \textit{tourist} vortex will not be seen if the beam waist is small and the TC is high. Taking this into account and applying the technique of counting vortices surrounded by $\mathcal{C}$ the ``strength staircase'' it is not going to be uniform for practical vortices and it is going to change from lower to higher transitions of TC. Figure \ref{fig3b} shows the ``staircase'' for three different values of $\omega_0$. This can also be read as changing the radius of a curve $\mathcal{C}$ where $0.3\rightarrow0.9$ could represent a change from shorter radius to a longer one. The problem in practical scenarios is that since the \textit{resident} enters before the \textit{tourist} there is a moment where the TC jumps to a higher value and then when the \textit{tourist} arrives (with unitary opposite charge) the TC returns to its previous value. This effect create the spikes that can be seen in every transition from odd to even. The form of the ``staircase'' differs from that of a regular one, that is one step per topological charge. Instead of that, with a higher value of the beam waist there is a tendency to odd values of TC with two fold jumps at even numbers of $\alpha$. 

\begin{figure}[ht!]
\centering\includegraphics[width=0.9\linewidth]{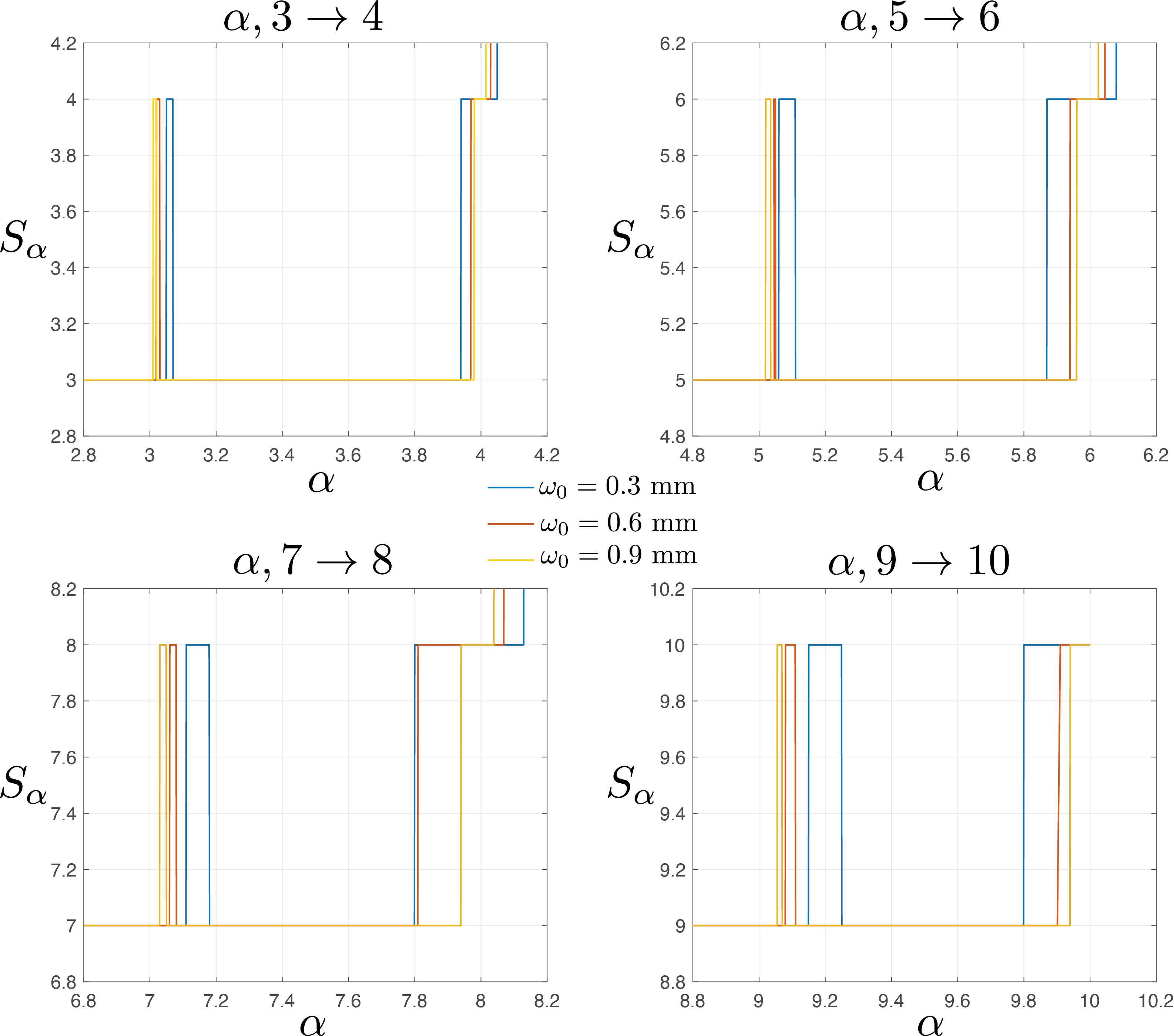}
\caption{Detail of the odd to even transitions comparing the different beam waists scenarios. This shows the tendency to shorten the gap for higher values of TC and also for smaller values of the beam waist.}
\label{fig3c}
\end{figure}

On the contrary, for lower values of $\omega_0$ the spikes widen themselves and the jumps at even numbers turn to a new step. 
In this case what would happen is that the stair will be regular but the jumps will occur at a non integer value. To explore this further, we show in Figure \ref{fig3c} the comparison of the odd to even transitions for the different beam waist scenarios. In general the width of the spike is not regular, but there is a clear movement towards the center for lower values of $\omega_0$. Also, the step that is produced at the even values of $\alpha$ moves away form the integer value. Those values are equivalent to the steepest curve of Figure \ref{fig1b} (a). And then for a transition with higher $\alpha$ when the \textit{tourist} curve doesn't exist the valley between the spike and the step will vanish. For higher values of $\omega_0$ the spike will be thinner and below the margin of error. This will lead to previous results seen in \cite{paper:Wen2019}.

\section{Conclusions}

In this work we have shown that the behavior of the vortices during the continuous transition between two integer values of the TC follow specific curves. Those curves depend not only on the TC but also on the beam waist. This poses the problem of estimating the vortex strength using either equation \ref{eq:000} or \ref{eq00} arriving to different results than the stated previously in literature. This discrepancy is only explained from the point of view of the real extension of the vortex beams and thus the use of the term \textit{practical} vortex. We have proved that even though the phase remains non integer at far field, the ``counting'' technique gives just integer results only dependent on the extension of the beam or equivalently the radius of the curve $\mathcal{C}$. Our analysis gives additional insight regarding the results obtained in practical situations and may be useful in applications such as optical communications, where the proper calculus of vortex strength is a key point.  

\paragraph{FUNDING} This work has been funded in part by project PID2019-110927RB-I00 financed by  MCIN/AEI/10.13039/501100011033 and project Prometeo/2020/029 financed by Generalitat Valenciana.
\paragraph{DISCLOSURES} The authors declare no conflicts of interest.


\begin{thebibliography}{99}

\newcommand{\enquote}[1]{``#1''}

\bibitem{paper:Allen1992}
L. Allen, M. W. Beijersberge, R. J. C. Spreeuw, and J. P. Woerdman, \enquote{Orbital angular momentum of light and the transformation of laguerre-Gaussian laser modes,} Phys. Rev. A, \textbf{45}, (1992) 8185--8189. \url{https://doi.org/10.1103/PhysRevA.45.8185}

\bibitem{paper:Beijersbergen1994}
M.W. Beijersbergen, R.P.C. Coerwinkel, M. Kristensen, J.P. Woerdman, \enquote{Helical-wavefront laser beams produced with a spiral phaseplate} Opt. Commun. \textbf{112} (1994) 321--327. 
\url{https://doi.org/10.1016/0030-4018(94)90638-6}

\bibitem{paper:Berry2004}
M. V. Berry, \enquote{Optical vortices evolving from helicoidal integer and fractional phase steps}, J. Opt. A: Pure Appl. Opt. \textbf{6}, (2004) 259--268.
\url{https://doi.org/10.1088/1464-4258/6/2/018}

\bibitem{paper:Basistiy2004}
I. V. Basistiy, V. A. Pas'ko, V. V. Slyusar, M. S. Soskin, and M. V.Vasnetsov, \enquote{Synthesis and analysis of optical vortices with fractional topological charges}, J. Opt. A: Pure Appl. Opt. \textbf{6}, (2004) S166--S169.
\url{https://doi.org/10.1088/1464-4258/6/5/003}

\bibitem{paper:Leach2004}
J. Leach, E. Yao, and M. J. Padgett, \enquote{Observation of the vortex structure of a non-integer
vortex beam}, New J. Phys. \textbf{6}, (2004) 71.
\url{https://doi.org/10.1088/1367-2630/6/1/071}

\bibitem{paper:WLee2004}
W. Lee, X.-C. Yuan, and K. Dholakia, \enquote{Experimental observation of optical vortex evolution in a Gaussian beam with an embedded fractional phase step}, Opt. Commun. \textbf{239}, (2004) 129--135.
\url{https://doi.org/10.1016/j.optcom.2004.05.035}

%% application

\bibitem{paper:STao2005}
S.~H.~Tao, X-C.~Yuan, J.~Lin, X.~Peng, and H.~B.~Niu, \enquote{Fractional optical vortex beam induced rotation of particles}, Opt. Express \textbf{13}, (2005) 7726--7731.
\url{https://doi.org/10.1364/OPEX.13.007726}

\bibitem{paper:Tkachenko2017}
G. Tkachenko, M. Chen, K. Dholakia, and M. Mazilu, \enquote{Is it possible to create a perfect fractional vortex beam?}, Optica \textbf{4}, (2017) 330--333 .
\url{https://doi.org/10.1364/OPTICA.4.000330}

\bibitem{paper:Bing2021}
Bing Gu, Yueqiu Hu, Xiaohe Zhang, Miao Li, Zhuqing Zhu, Guanghao Rui, Jun He, and Yiping Cui, \enquote{Angular momentum separation in focused fractional vector beams for optical manipulation,} Opt. Express \textbf{29}, (2021) 14705-14719.
\url{https://doi.org/10.1364/OE.423357}

\bibitem{paper:Liu2021}
Hongyan Liu; Yu Wang; Jianqiu Wang; Kang Liu; Hongqiang Wang
\enquote{Electromagnetic Vortex Enhanced Imaging Using Fractional OAM Beams}
IEEE Antenn. Wirel. Pr. \textbf{20}, 6 (2021) 948--952.
\url{https://doi.org/10.1109/LAWP.2021.3067914}

\bibitem{paper:shuhuili2017}
Shuhui Li and Jian Wang, \enquote{Experimental demonstration of optical interconnects exploiting orbital angular momentum array}, Opt. Express \textbf{25} (2017) 21537--21547.
\url{https://doi.org/10.1364/OE.25.021537}

\bibitem{paper:Duodeng2021}
Duo Deng, Yanhao Chu, Yi Liu, Yan Li, Yanhua Han, \enquote{Measurement of multiplexed fractional vortices with integer mode interval}, Results Phys. \textbf{29} (2021) 104699.
\url{https://doi.org/10.1016/j.rinp.2021.104699}

\bibitem{paper:Kuang2020}
Kuang Zhang, Yueyi Yuan, Xumin Ding, Haoyu Li, Badreddine Ratni, Qun Wu, Jian Liu, Shah Nawaz Burokur, and Jiubin Tan
\enquote{Polarization-Engineered Noninterleaved Metasurface for Integer and Fractional Orbital Angular Momentum Multiplexing}
Laser Photonics Rev. \textbf{15}, (2020) 2000351.
\url{https://doi.org/10.1002/lpor.202000351}

\bibitem{paper:Alperin2016}
S. N. Alperin, R. D. Niederriter, J. T. Gopinath, and M. E. Siemens, \enquote{Quantitative measurement of the orbital angular momentum of light with a single, stationary lens,} Opt. Lett. \textbf{41}, (2016) 5019-5022.
\url{https://doi.org/10.1364/OL.41.005019}

\bibitem{paper:Gbur2016}
G. Gbur, \enquote{Fractional vortex Hilbert’s Hotel}, Optica \textbf{3}, (2016) 222--225 .
\url{https://doi.org/10.1364/OPTICA.3.000222}

\bibitem{paper:Matta2020}
S. Matta, P. Vayalamkuzhi, N. K. Viswanathan, \enquote{Study of fractional optical vortex beam in the near-field } Opt. Commun. \textbf{475} (2020) 126268.
\url{https://doi.org/10.1016/j.optcom.2020.126268}

\bibitem{paper:Duo2019}
Duo Deng, Muchun Lin, Yan Li, and Hua Zhao \enquote{Precision Measurement of Fractional Orbital Angular Momentum}, Phys. Rev. Applied \textbf{12}, (2019) 014048.
\url{https://doi.org/10.1103/PhysRevApplied.12.014048}

\bibitem{paper:Hosseini2020}
S. Hosseini-Saber, E. A. Akhlaghi, and A. Saber, \enquote{Diffractometry-based vortex beams fractional topological charge measurement,} Opt. Lett. \textbf{45}, (2020) 3478--3481.
\url{https://doi.org/10.1364/OL.395440}

\bibitem{paper:Guoxuan2021}
Guoxuan Zhu, Zhao liu , Cailing Fu, Shen Liu , Zhiyong Bai , and Yiping Wang ,
\enquote{High-Precise Fractional Orbital Angular Momentum Probing With a Fiber Grating Tip}
J. Light. Technol. \textbf{39} 6, (2021) 1867--1872.
\url{https://doi.org/10.1109/JLT.2020.3042602}

\bibitem{paper:Alperin2017}
S. N. Alperin and M. E. Siemens, \enquote{Angular Momentum of Topologically Structured Darkness}, Phys. Rev. Lett. \textbf{119}, (2017) 203902.
\url{https://doi.org/10.1103/PhysRevLett.119.203902}

\bibitem{paper:Wen2019}
J. Wen, L.-G. Wang, X. Yang, J. Zhang, and S.-Y. Zhu, \enquote{Vortex strength and beam propagation factor of fractional vortex beams}, Opt. Express \textbf{27}, (2019) 5893--5904.
\url{https://doi.org/10.1364/OE.27.005893}

\bibitem{paper:Jesus-Silva2012}
A. J. Jesus-Silva, E. J. S. Fonseca, and J. M. Hickmann, \enquote{Study of the birth of a vortex at Fraunhofer zone}, Opt. Lett.\textbf{37}, (2012) 4552--4554.
\url{https://doi.org/10.1364/OL.37.004552}






















\end{thebibliography}
\end{document}